\documentclass[amsmath,amssymb,superscriptaddress,twocolumn]{revtex4-1}

\newcommand{\be}{\begin{eqnarray}}
\newcommand{\ee}{\end{eqnarray}}
\def\({\left(}
\def\){\right)}
\def\[{\left[}
\def\]{\right]}

\newcommand{\braket}[1]{\langle #1 \rangle}
\newcommand{\bra}[1]{\langle #1 |}
\newcommand{\ket}[1]{| #1 \rangle}

\newcommand{\sla}[1]{\rlap{\kern .15em /}#1}

\def\su{\mathfrak{su}}
\def\so{\mathfrak{so}}
\def \sig#1#2 {\sigma_{#1} \otimes \sigma_{#2}}

\begin{document}

\title{Non-Adiabatic Universal Holonomic Quantum Gates Based on Abelian Holonomies}

\author{Utkan G\"ung\"ord\"u}
\affiliation{Research Center for Quantum Computing, Interdisciplinary Graduate School of Science and Engineering, Kinki University, 3-4-1 Kowakae, Higashi-Osaka, Osaka 577-8502, Japan}
\email{utkan@alice.math.kindai.ac.jp}

\author{Yidun Wan}
\affiliation{Department of Applied Physics, Graduate School of Engineering, University of Tokyo, Hongo 7-3-1, Bunkyo-ku, Tokyo 113-8656, Japan}
\affiliation{Perimeter Institute for Theoretical Physics, 31 Caroline St N, Waterloo, ON N2L 2Y5, Canada}
\email{ywan@meso.t.u-tokyo.ac.jp}

\author{Mikio Nakahara}
\affiliation{Research Center for Quantum Computing, Interdisciplinary Graduate School of Science and Engineering, Kinki University, 3-4-1 Kowakae, Higashi-Osaka, Osaka 577-8502, Japan}
\affiliation{Department of Physics, Kinki University, 3-4-1 Kowakae, Higashi-Osaka, Osaka 577-8502, Japan}
\email{nakahara@math.kindai.ac.jp}

% \date{\today}
\begin{abstract}
We implement a non-adiabatic universal set of holonomic quantum gates based on abelian holonomies using dynamical invariants, by Lie-algebraic methods. Unlike previous implementations, presented scheme does not rely on secondary methods such as double-loop or spin-echo and avoids associated experimental difficulties. It turns out that such gates exist purely in the non-adiabatic regime for these systems. 
\end{abstract}

%%% Keywords are not needed any longer. %%%
%%%\kword{keyword1, keyword2, keyword3, \ldots}
%%%

\maketitle

\section{Introduction}

A quantum computer is expected to outperform a classical digital
computer in some of computationally hard tasks \cite{Nakahara2008,Nielsen2000,Jones2012}. The prime factorization, for example,
is widely believed to be an NP problem for a classical computer but has merely polynomial complexity (BQP)
with a quantum computer and Shor's algorithm.

It is well known that any classical logic operation may be
realized by a collection of the NAND gate. The corresponding
``universality theorem'' in quantum circuits is due to Barenco {\it et al.}
\cite{Barenco1995}. The theorem claims that any unitary gate
can be decomposed into one-qubit (i.e., SU(2)) gates and CNOT
gates. In other words, the set of one-qubit gates and the CNOT
gate are universal in arbitrary gate implementations. In many physical
systems, implementing a one-qubit gate is often not difficult; it may be realized by the Rabi oscillation or the Raman
transition, for example. In contrast, implementing the
CNOT gate can be challenging and its realization is sometimes regarded
as a milestone for a physical system to be a true candidate of
a working quantum computer \cite{DiVincenzo2000}.
Later, it turned out that any SU(4) gate, which entangles
a tensor product state, may serve as an element of a universal
set of quantum gates with the set of one-qubit gates
\cite{Barenco1995a,Lloyd1995}. Important exceptions of two-qubit
gates that are excluded are the SWAP gate and
the ``local gates'' $\text{SU}(2)\otimes \text{SU}(2)$.

There are many nontrivial quantum algorithms, such as the Deutsch Algorithm,
the Grover algorithm and the Bernstein-Vazirani algorithm, just
to name a few, which can be demonstrated with a two-qubit system
and an $\text{SU}(4)$ gate. Execution of these algorithms with a speed
beyond the adiabatic limit shows the promising future of
a realization of quantum computing.

In spite of all expectations, however, implementation of a working quantum
computer is yet to be realized. One of the hardest obstacles
against its physical realization is decoherence. 
A quantum system is subject to environment and interaction
between them degrades the purity of the quantum system
to be used as a hardware of a quantum computer.

Another obstruction against the realization of a working
quantum computer is the gate fidelity. Quantum gates
used for quantum computation are essentially analogue
gates. Implementation of high-precision gates is an urgent
subject to be solved. Composite gates originally proposed
in NMR measurement are one of such strategies although
composite gates require many elementary gates and
hence longer execution time \cite{Ichikawa2012,Ichikawa2013}.

Non-adiabatic geometric quantum gates (GQG)
are expected to solve these problems simultaneously.
Geometric gates make use of the holonomy associated with
an underlying fiber bundle structure of a quantum system \cite{Zanardi1999}. 
If one employs the Berry phase and the Wilczek-Zee holonomy
for implementation of quantum gates, however, the gate execution time
is strongly limited by adiabaticity, which may exceed decoherence times. Therefore it is desirable to
implement gates by employing non-adiabatic control.

Non-adiabatic implementations based on non-abelian holonomies have recently been reported in \cite{Solinas2003,Sjoqvist2012,Xu2012,Feng2013a,Abdumalikov2013}.
A non-adiabatic geometric quantum computation (GQC) scheme based on abelian holonomies was proposed by Zhu and Wang \cite{Zhu2002, Zhu2002erratum, Zhu2003} for NMR and Josephson charge qubits.
The implementation is based on the Aharonov-Anandan (AA) 
phase $\gamma^g$ associated
with cyclic changes of the states \cite{Aharonov1987,Anandan1990}. The total phase associated with the evolution of a cyclic vector $\ket{\phi(t)}$
separates into two parts, the dynamical phase
$\gamma^d = -\int \bra{\phi(t)}H(t)\ket{\phi(t)} dt$
and the geometric phase $\gamma^g = \oint i \bra{\phi(t)}
d\ket{\phi(t)}$ (we use natural units where $\hbar = 1$). Furthermore, $\ket{\phi(t)}$ is an eigenvector
of the time-evolution operator. In previous works \cite{Zhu2002, Zhu2002erratum, Zhu2003}, authors employed two loops in the
control parameter space so that the dynamical phases
from the two loops cancel with each other while the geometric
phases accumulate (the double-loop method). 
Note, however, that using a second loop
is far from optimal and, moreover, experimental feasibility
may be challenging. In the NMR context,
it involves rotation of a heavy superconducting magnet and
hence is not practical.

Later, Ota {\it et al.} replaced the second loop by 
a pair of $\pi$-pulse, which is known as the spin-echo
technique in the NMR community and implemented the proposal
using an NMR quantum computer \cite{Ota2009}. It still requires the extra pair
of $\pi$-pulses and may not be the optimal implementation of 
quantum gates. Furthermore, this technique is applicable only
to one-qubit gates and implementation of two-qubit gates is
still an open question in this framework.
It should be noted that Ota {\it et al.}  report that the two-qubit gate implementation given by Zhu and Wang \cite{Zhu2003} may be erroneous.

Here, we report a new implementation of 
non-adiabatic geometric gates based on cyclic states with vanishing dynamical phases. The end result is a time
evolution whose only measurable effects are due to holonomies.
Our approach is Lie-algebraic and does not
require double-loop nor spin-echo technique, making it suitable for practical applications.
As it turns out, such a scheme exists only in the non-adiabatic regime.
In our implementation, the gate execution can be made fast by choosing a short period.

This paper is organized as follows. In the next section, we describe dynamical invariants and how they can be used for GQC and in the following sections, where we implement a universal set of geometric quantum gates based on practical Hamiltonians.
Sections \ref{sec:singlequbit} and \ref{sec:twoqubit} explicitly construct one- and two-qubit gates respectively. Section \ref{sec:Conc} concludes the paper.
\section{Geometric Quantum Gates and Dynamical Invariants}
% \label{sec:lri}
Let us give a brief review of dynamical invariants \cite{Lewis1969} and describe their relation with GQGs. A dynamical invariant $I$ is an explicitly time-dependent observable whose expectation value is yet a constant, and thus obeys the Liouville-von Neumann equation
\be
0 = \partial I/\partial t + i[H, I]
\label{eq:DIequation}
\ee
where $H$ is the Hamiltonian of the system.
Its eigenvectors $\{\ket{\phi_n(t)}\}$ are related to the solutions of the Schr\"odinger equation by a local gauge transformation 
$\ket{\psi_n(t)} = e^{i \alpha_n(t)}\ket{\phi_n(t)}$
where the Lewis-Riesenfeld phase $\alpha_n(t)$ is given by (see Appendix \ref{sec:LRphase})
\be
\int_0^t \bra{ \phi_n(s)} \left(i\frac{d}{d s} - H(s) \right) \ket{\phi_n(s)} ds. 
\label{eq:LRphase}
\ee 
The Aharonov-Anandan phase, which has recently been observed directly in an electronic spin \cite{Nagasawa2013}, is the U(1) fiber bundle structure that manifests itself in cyclic evolutions in the projective Hilbert space, in the form $\ket{\psi_n(T)} = e^{i\gamma_n^g} \ket{\psi_n(0)}$ \cite{Aharonov1987,Anandan1990}. Let us take a 
single-valued, closed representative of this curve $\ket{\phi_n(t)}$ $(0 \leq t \leq T)$ defined through a local U(1) gauge transformation $e^{i \alpha_n(t)}\ket{\phi_n(t)} = \ket{\psi_n(t)}$, and require $\ket{\phi_n(T)} = \ket{\phi_n(0)}$. Using the Schr\"odinger equation, $\alpha_n(t)$ is found to obey Eq. (\ref{eq:LRphase}), thus the Lewis-Riesenfeld phase for a cyclic evolution is closely related to the Aharonov-Anandan phase (see Appendix \ref{sec:AAphase}). The phase is made of 
two terms. The holonomy (or the geometric phase)
\be
\gamma_n^g = \oint i \bra{\phi_n(t)} d\ket{ \phi_n (t)},
\ee
determined by the one-form connection, depends only on the path taken in the projective Hilbert space, and is invariant under time-reparametrization $t \to \tau(t)$. The gauge-invariant term
\be
\gamma_n^d = -\int_0^T \bra{ \phi_n(t)} H(t) \ket{\phi_n(t)}dt
\ee
is called dynamical phase.
% The evolution becomes purely geometric when the dynamical phase vanishes, enabling the construction of geometric quantum gates \cite{Ichikawa2012,Teo2005}.

The unitary time-evolution operator can be written in terms of the eigenvectors of $I$  and $\alpha_n(t)$ as
\be
U(t;0) = \sum_n e^{i \alpha_n(t)}\ket{\phi_n(t)} \bra{\phi_n(0)}.
\label{eq:time-evolution}
\ee
When the dynamical phases vanish, the time-evolution is dictated only by the geometric terms at $t=T$, and the resulting operation becomes a geometric quantum gate \cite{Ichikawa2012,Sjoqvist2013,Teo2005}.

A corollary of Eq. (\ref{eq:time-evolution}) is that the eigenstates of $I$ evolve in the simple form 
\be
U(t;0)|\phi_n(0)\rangle=e^{i \alpha_n(t)}|\phi_n(t)\rangle,
\ee
which alludes the adiabatic time-evolution of a quantum state. However, this passage is transitionless in the eigenbasis $\{ \ket{\phi_n(t)} \}$, and is not restricted by adiabaticity condition.
A major problem in schemes based on adiabatic evolution is that in many cases the evolution is so slow that the system may start decohering and operational times are bounded from below. In contrast, non-adiabatic schemes can be made fast, making them better candidates for GQC.

\section{One-Qubit Gates}
\label{sec:singlequbit}
To realize single-qubit operations, we start with the time-dependent Hamiltonian
\be
H = \frac{1}{2}(\Omega \cos\omega t \sigma_x + \Omega \sin\omega t \sigma_y + \Delta \sigma_z),
\label{eq:H}
\ee
which can be used to describe a semi-classical dipole-electric field interaction, spin-magnetic field interaction or a Josephson charge qubit \cite{Vartiainen2004,Zhu2002, Zhu2002erratum} among other things.

The corresponding dynamical invariant for this system is 
\be\label{di1}
I = \Omega \cos\omega t \sigma_x + \Omega \sin\omega t \sigma_y + (\Delta-\omega) \sigma_z,
\ee
and its eigenvalues and eigenvectors are
\be
\pm\lambda, \quad \ket{\phi_\pm(t)} = 
\begin{pmatrix}
e^{-i \omega t}\cos \theta_{\pm}\\
\sin \theta_{\pm}
\end{pmatrix}
\ee
where
\be
\cos \theta_{\pm}  = \xi_{\pm}/\sqrt{1+\xi_{\pm}^2}, \qquad \sin \theta_{\pm} = 1/\sqrt{1+\xi_{\pm}^2} \nonumber\\
\xi_\pm = [(\Delta-\omega) \pm \lambda]/\Omega, \lambda = \sqrt{\Omega^2+(\Delta-\omega)^2}.
\ee
By direct evaluation,
the corresponding Lewis-Riesenfeld phases of these vectors are found to be $\alpha_\pm(t) = (\omega \mp \lambda) t/2$.
$H$, $I$ and $\ket{\phi_\pm(t)}$ are cyclic in time with period $T=2\pi/\omega$ and accordingly, after one cycle of evolution, solutions of the Schr\"odinger equation $\ket{\psi_\pm(t)}$ pick up the phases $\ket{\psi_\pm(T)}=e^{i \alpha_\pm(T)}\ket{\psi_\pm(0)}$.

To implement a GQG, one needs to have a vanishing dynamical phase. It is possible to cancel the dynamical phases by using a second loop \cite{Zhu2002, Zhu2002erratum, Zhu2003, Ekert2000} but this comes at the cost of doubling the number of gates to realize quantum algorithms and an increased total operational time, and can cause difficulties in realization \cite{Ota2009}. The condition $\int_0^T \bra{\phi_\pm (t)} H(t) \ket{\phi_\pm (t)}dt = 0$ yields
\be
\Omega^2+\Delta(\Delta-\omega) = 0,
\label{eq:onequbitcondition}
\ee
and this in turn means $\Delta \in (0,\omega)$  and $\Omega^2 \approx \Delta \omega$. Clearly, $\Omega$ should be non-zero given that we would like to implement non-trivial universal quantum gates. These conditions cannot be met in the adiabatic regime ($\delta E = \sqrt{\Omega^2 + \Delta^2} \gg 1/T$ where $\delta E$ is the gap between the energy eigenvalues of the Hamiltonian), hence we conclude that such a gate is purely a result of non-adiabatic effects.

The resulting SU(2) gate operation is
\be
U_\beta(T) = -e^{i\pi \sin\beta \left[-\cos\beta\sigma_x+\sin\beta\sigma_z\right ]},
\ee
where we defined  $\cos^2\beta=\Delta/\omega$, $\beta \in (0,\pi/2)$, in the computational basis which has a single free-parameter $\beta$ and can produce two non-commuting one-qubit operations which are required to realize universal quantum computation \cite{Lloyd1995}.

We remark that the well-known result Eq. (\ref{eq:onequbitcondition}) has also been mentioned by Zhu and Wang \cite{Zhu2003}, using an alternative derivation. The major advantage of our dynamical invariant based approach is its extensibility that goes beyond two-level systems. Even though the set of independent operators generating the one-qubit Hamiltonian is the fundamental representation of $\su(2)$, the approach we take here is algebraic in nature and can be applied to three-level systems such as neutral atoms in cavity QED \cite{Recati2002} due to the isomorphism $\su(2) \cong \so(3)$ and through possible embeddings of $\su(2)$ into $\su(n)$, $n\geq 3$, since the set of cyclic vectors $\{\ket{\phi_n}\}$ can be obtained by using the dynamical invariant of the $\su(2)$ problem. Clearly, the conditions for making the dynamical phases vanish will be different, but given the dynamical invariant, they can be obtained straightforwardly. In the next section, we demonstrate this for the embedding $\su(2) \subset 
\su(4)$.

\section{Two-Qubit Gate}
\label{sec:twoqubit}
In addition to one-qubit operations, a gate that is capable of creating entanglement between two qubits is required to achieve universality. We realize this by introducing a second qubit which is coupled to the original control qubit 
through the Ising interaction $\sigma_z \otimes \sigma_z$ with limited control,
\be
H' = \frac{1}{2} J \sigma_z \otimes \sigma_z +  \openone \otimes H_c + q(t) \sigma_z \otimes \openone,
\ee
where
\be
H_c = (\Omega \cos \omega t \sigma_x + \Omega \sin \omega t \sigma_y + D \sigma_z)/2
\ee
and $\openone$ is the $2 \times 2$ identity matrix. This kind of Hamiltonian appears in the study of Josephson charge qubits or liquid-state NMR \cite{Nakahara2008,Nielsen2000,Zhu2002}. Two-qubit Hamiltonians are elements of $\su(4)$ in general (the identity term can always be dropped as it only contributes a global phase, which can be restored at any time)
which makes an analytic approach difficult. However, the problem can be systematically reduced when the Hamiltonian is restricted to a subalgebra of $\su(4)$ \cite{Gungordu2012a}. In this particular case, the generating set of $H'$ spans the subalgebra $\su(2) \oplus \su(2) \oplus \mathfrak u(1)$, which is
\be
\sqcup_{s \in \{+,-\} } \{G^s_x, G^s_y, G^s_z \} \sqcup \{\sigma_z \otimes \openone\},
\ee
where $G^s_i = (\openone + s \sigma_z)/2 \otimes \sigma_i$, thus the problem can be treated as two non-interacting logical qubits evolving independently in this artificial basis.

Note, however, that the basis vectors $\{G_i^s\}$ create entanglement
between two physical qubits and are able to implement non-trivial SU(4) gates.
Such embedding of $\mathfrak{su}(2)$ subalgebra in $\mathfrak{su}(4)$
has been reported \cite{Ichikawa2013} in the context of implementation of
high precision SU(4) gates in the presence of coupling strength
errors, where SU(2) gates robust against pulse length error 
were mapped to SU(4) gates.

Due to the algebraic structure, the time-dependent $\mathfrak u(1)$ term $q(t) \sigma_z \otimes \openone$ is incorporated without complicating the problem. The outcome is a local U(1) term
\be
U_q' = \exp\left(-i \sigma_z \otimes \openone \int_0^t q(s) ds\right)
\ee
in the gate operation, however, it turns out that the main results below are unaffected by the choice for $q(t)$: it does not appear in the invariant due to the commutator in the Liouville-von Neumann equation Eq. (\ref{eq:DIequation}), and it will not affect the entangling part of the gate since it is a local operation. However, it can be considered as an additional degree of freedom, providing with a free one-qubit operation.

The algebra decomposition allows us to write $H'$ as a sum of three commuting parts 
\be
H'_\pm =( \Omega \cos\omega t G^\pm_x +  \Omega \sin\omega t G^\pm_y + \Delta_\pm G^\pm_z )/2
\label{eq:Hpm}
\ee
and $q(t) \sigma_z \otimes \openone$, with $\Delta_\pm = D \pm J$. Since $H'_\pm$ in Eq. (\ref{eq:Hpm}) have the same functional form as the single-qubit Hamiltonian $H$ in Eq. (\ref{eq:H}), by replacing $\Delta \to \Delta_\pm$ and $\sigma_i \to G_i^\pm$ in the one-qubit invariant $I$, we can obtain their corresponding dynamical invariants as 
\be\label{di2}
I'_\pm = \Omega \cos\omega t G^\pm_x + \Omega \sin\omega t G^\pm_y + (\Delta_\pm - \omega) G^\pm_z.
\ee
$H'_\pm$ commute with each other at all times and as a result there will be two commuting SU(2) time-evolution operators $U'_\pm$ representing the evolution of the two separate subspaces and the total gate operation can be written as $U'=U_q' U'_+ U'_-$.

For evaluation of the geometric phases corresponding to the solutions of the Schr\"odinger equation, we prefer to use a simpler dynamical invariant $I'=I_+' + I_-'$ whose eigenvectors $\ket{\phi_\pm^+(t)}, \ket{\phi_\pm^-(t)}$ are given as
\be
\begin{pmatrix}
e^{-i \omega t} \cos \theta_{\pm}^+ \\
\sin \theta_{\pm}^+\\
0 \\
0
\end{pmatrix},
\begin{pmatrix}
0 \\
0 \\
e^{-i \omega t} \cos \theta_{\pm}^- \\
\sin \theta_{\pm}^-
\end{pmatrix},
\ee
where
\be
\cos \theta_{\pm}^{s} = \xi_{\pm}^{s}/\sqrt{1+{\xi_{\pm}^{s}}^2}, \qquad \sin \theta_{\pm}^{s} = 1/\sqrt{1+{\xi_{\pm}^{s}}^2}\nonumber\\
\xi_{\pm}^{s} = [(\Delta_{s}-\omega) {\pm} \lambda_{s}]/\Omega, \qquad \lambda_\pm = \sqrt{\Omega^2+(\Delta_\pm-\omega)^2}.
\ee
Their corresponding eigenvalues are $\pm\lambda_+$ and $\pm\lambda_-$, and the Lewis-Riesenfeld phases are $(\omega \mp \lambda_+) t/2$ and $(\omega \mp \lambda_-) t/2$ respectively.
The requirement for all dynamical phases to vanish simultaneously is found to be equivalent to requiring
\be
D = \omega/2, \qquad \Omega = \pm \sqrt{(\omega/2)^2 - J^2},
\ee
and hence $\omega/2 > |J|$. This constraint has also been mentioned by Zhu and Wang \cite{Zhu2003} without being used, where they discuss the special case $q(t)=0$ using a different method; the reported non-trivial two-qubit gate implementations are incompatible with Eq. (\ref{eq:twoqubit-constraint}) and rely on a second-loop to get rid of the dynamical phases instead.

Similarly to the one-qubit gate, we find that it is not possible to satisfy these conditions in the adiabatic regime.

Universality can be achieved by requiring a two-qubit quantum gate that is capable of transforming tensor product states into maximally entangled states. A two-qubit quantum gate is local if it belongs to the subgroup $\text{SU}(2) \otimes \text{SU}(2)$ generated by $\text{span}(\sqcup_i \{\openone \otimes \sigma_i\}) \oplus \text{span}(\sqcup_i \{\sigma_i \otimes \openone\})$, which typically corresponds to the physical spin or polarization eigenstates. A non-local gate is called a perfect entangler if it can produce a maximally entangled state from a tensor product state \cite{Makhlin2002, Zhang2003, Balakrishnan2011}, such as CNOT gate. In contrast, SWAP gate is a typical non-perfect entangler \cite{Makhlin2002, Zhang2003}.
The Hamiltonian $H'$ is capable of creating a GQG that is as well a perfect entangler for a suitable choice of parameters which can be determined by operator Schmidt decomposition \cite{Nielsen2003}. The non-vanishing singular values $D_\pm$ of the matrix $C_{i j} = \text{tr}\left( U' \sigma_{i} \otimes \sigma_{j}  \right)/4$ are found to be
\be
\sqrt{\frac{1}{2} \pm  \left(\frac{1}{2}\cos \pi a_+ \cos \pi a_- + a_+ a_- \sin \pi a_+ \sin \pi a_- \right)} \nonumber \\
\ee
where $a_\pm = \sqrt{1/2\pm J/\omega}$. The result $\text{rank}(C)=2$ implies that such a configuration may implement a CNOT gate, but not a SWAP gate whose
rank is 4 \cite{Zhang2003}. For this gate to become a perfect entangler, $D_\pm$ must match the corresponding singular values of CNOT, which are $D_\pm^{\text{CNOT}} = \sqrt{1/2}$. Hence,
\be
\cot \pi a_+ \cot \pi a_- + 2 a_+ a_- = 0
\label{eq:twoqubit-constraint}
\ee
must be satisfied, which has physically acceptable solutions
\be
r = J/\omega \approx \pm 0.3187.
\ee
The operational time $T$ is $2 \pi |r| /J$. The entangling part of the gate, which can be found by methods such as Cartan-decomposition \cite{Nakahara2008,Helgason2001}, is $e^{i \pi \sigma_y \otimes \sigma_y /4}$.
We note that implementation of such a gate requires non-vanishing transverse fields ($\Omega \neq 0$) cf. \cite{Xiang-Bin2001,Xiang-Bin2002}.

Another common Ising-type coupling is $\sigma_x \otimes \sigma_x$ term. Using the same framework, it is possible to handle this problem with minor modifications. The Hamiltonian still belongs to the $\su(2) \oplus \su(2) \oplus \mathfrak u(1)$ subalgebra, but this time with generators
\be
\sqcup_{s \in \{+,-\} } \{G'^s_x, G'^s_y, G'^s_z \} \sqcup \{\sigma_x \otimes \openone\},
\ee
where $G'^s_i = (\openone + s \sigma_x)/2 \otimes \sigma_i$. Using the corresponding dynamical invariants \cite{Gungordu2012a}
\be
\Omega \cos\omega t G'^\pm_x + \Omega \sin\omega t G'^\pm_y + (\Delta_\pm - \omega) G'^\pm_z,
\ee
one can repeat the straightforward but tedious analysis (obtaining eigenvectors of the dynamical invariant, enforcing that dynamical phases vanish and that the gate is local unitarily equivalent to a non-trivial gate) and obtain a non-trivial two-qubit GQG.

\section{Conclusion and Outlook}\label{sec:Conc}
In this work, we have implemented a universal set of non-adiabatic geometric quantum gates based on abelian holonomies by making use of the dynamical invariants (\ref{di1}) and 
(\ref{di2}). Our implementation does not require secondary techniques such as double-loop or spin-echo technique and is expected to be physically straightforward on NMR.
We remark that the Lie-algebraic method employed above can be used to construct non-adiabatic quantum gates based on non-abelian holonomies \cite{Anandan1988,Bohm1994} exhibiting a richer geometrical structure \cite{Bohm2003}, which we leave for future work.

\section*{Acknowledgments}
\begin{acknowledgments}
UG and MN are grateful to JSPS (Japan Society for the Promotion of Science)
for partial support from Grant-in-Aid for Scientific Research (Grant
No. 24320008). MN also thanks JSPS for Grant-in-Aid
for Scientific Research (Grant No. 23540470).
UG acknowledges the financial support of the MEXT (Ministry of
Education, Culture, Sports, Science and Technology)
Scholarship for foreign students.
YW would like to thank Profs. Seigo Tarucha and Rod Van Meter.
YW is grateful for the hospitality of the Perimeter Institute for Theoretical Physics.
This research was supported in part by Perimeter Institute for Theoretical Physics. Research at Perimeter Institute is supported by the Government of Canada through Industry Canada and by the Province of Ontario through the Ministry of Economic Development \& Innovation.
\end{acknowledgments}

\appendix
\section{Lewis-Riesenfeld Phase}
\label{sec:LRphase}
$I$ is a conserved observable by definition and as a result its spectral decomposition is $I = \sum_n \lambda_n \ket{\phi_n(t)} \bra{\phi_n(t)}$ with time-independent eigenvalues $\lambda_n$. It follows that $\lambda_n \ket{\dot \phi_n(t)} = \partial_t I \ket{\phi_n(t)} + I \ket{\dot \phi_n(t)}$. Using Eq. (\ref{eq:DIequation}) to replace $\partial_t I$ term with the commutator and multiplying from the left by $\bra{\phi_m(t)}$, we obtain
\be(\lambda_m - \lambda_n)(\bra{\phi_m(t)}H\ket{\phi_n(t)} - i \braket{\phi_m(t) | \dot\phi_n(t)}) = 0. \nonumber \\
\ee
For $m \neq k$, this means $\bra{\phi_m(t)}H\ket{\phi_n(t)} - i \braket{\phi_m(t) | \dot\phi_n(t)} = 0$.
Clearly, it does not hold for $m = n$ in general. Let us assume that $\ket{\psi_n(t)} = e^{i \alpha_n(t)} \ket{\phi_n(t)}$ is a solution of the time-dependent Schr\"odinger equation, which translates to $-\dot \alpha_n(t) e^{i \alpha_n(t)} \ket{\phi_n(t)} + e^{i \alpha_n(t)} \ket{\dot \phi_n(t)} = H e^{i \alpha_n(t)} \ket{\phi_n(t)}$. This would only hold if
\be
\dot\alpha_n(t) = \bra{\phi_n(t)}(i d/dt -H)\ket{\phi_n(t)}
\ee
is satisfied, which is equivalent to Eq. (\ref{eq:LRphase}).

\section{Aharonov-Anandan Phase}
\label{sec:AAphase}
Let us assume that $\ket{\psi_n(t)}$ is a cyclic state in the Hilbert space such that $\ket{\psi_n(T)} = e^{i \Theta_n}\ket{\psi_n(0)}$.
We take a projection of this curve, a single-valued closed representative, $\ket{\phi_n(t)}$ $(0 \leq t \leq T)$, defined through a local U(1) gauge transformation $e^{i \theta_n(t)}\ket{\phi_n(t)} = \ket{\psi_n(t)}$, and require $\ket{\phi_n(T)} = \ket{\phi_n(0)}$ (thus, $\ket{\phi_n(t)}$ lives
in the projective Hilbert space). As a result, $\Theta_n = \theta_n(T) - \theta_n(0)$. Using the Schr\"odinger equation,
\be
\dot \theta_n(t) = i \braket{\phi_n(t) | \dot \phi_n(t)} - \bra{\phi_n(t)}H\ket{\phi_n(t)},
\ee
we find that $\theta_n(t)$ obeys Eq. (\ref{eq:LRphase}), and hence $\theta_n(t) \equiv \alpha_n(t)$. The second term in Eq. (\ref{eq:LRphase}) that is a functional of $H$ is called the dynamical phase, and the first term (the holonomy) is called the Aharonov-Anandan phase.

\bibliographystyle{jpsj}
\bibliography{lri,extra}

% \begin{thebibliography}{9}
% \bibitem{jpsj} The abbreviation for JPSJ must be ``J. Phys. Soc. Jpn." \note{in the reference list}.
% \bibitem{instructions} More abbreviations of journal titles are listed in ``Instructions for Preparation of Manuscript".
% \end{thebibliography}

\end{document}